\documentclass{iau} 
\usepackage{graphicx}
\usepackage{natbib}

\newcommand{\araa}{\textit{ARA\&A}} 
\newcommand{\mnras}{\textit{MNRAS}} 
\newcommand{\aap}{\textit{A\&A}} 
\newcommand{\aaps}{\textit{A\&AS}} 
\newcommand{\apj}{\textit{ApJ}} 
\newcommand{\zap}{\textit{Z. Astrophys.}} 
\newcommand{\jas}{\textit{J. Atmos. Sci.}} 

\title[Progenitors of Core-Collapse Supernovae] 
{Progenitors of Core-Collapse Supernovae}

\author[R. Hirschi et al]   
{R. Hirschi$^{1,2}$,
   D. Arnett$^3$, A. Cristini$^1$, C. Georgy$^4$, C. Meakin$^{3,5}$ \and I. Walkington$^1$}

\affiliation{$^1$ Astrophysics Group, Keele University, Lennard-Jones Laboratories, Keele, ST5 5BG, UK \\ email: {\tt r.hirschi@keele.ac.uk} \\[\affilskip]
$^2$ Kavli IPMU (WPI), The University of Tokyo, Kashiwa, Chiba 277-8583, Japan \\
$^3$ Department of Astronomy, University of Arizona, Tucson, AZ 85721, USA\\
$^4$ Geneva Observatory, University of Geneva, Ch. Maillettes 51, 1290 Versoix, Switzerland\\
$^5$ Karagozian \& Case, Inc., 700 N. Brand Blvd. Suite 700, Glendale, CA, 91203, USA
}

\pubyear{2017}
\volume{331}  
\jname{SN 1987A, 30 years later}
\editors{A. Marcowith, G. Dubner, A. Ray, A. Bykov, \& M. Renaud, eds.}
\begin{document}

\maketitle

\begin{abstract}
Massive stars have a strong impact on their surroundings, in particular when they produce a core-collapse supernova at the end of their evolution. In these proceedings, we review the general evolution of massive stars and their properties at collapse as well as the transition between massive and intermediate-mass stars. We also summarise the effects of metallicity and rotation. We then discuss some of the major uncertainties in the modelling of massive stars, with a particular emphasis on the treatment of convection in 1D stellar evolution codes. Finally, we present new 3D hydrodynamic simulations of convection in carbon burning and list key points to take from 3D hydrodynamic studies for the development of new prescriptions for convective boundary mixing in 1D stellar
evolution codes.

\end{abstract}

\firstsection 


\section{Introduction}
\label{intro}  
The progenitors of (iron) core-collapse supernovae (CCSNe) are by definition massive stars. Indeed, massive stars are stars massive enough to go through all the burning stages from hydrogen to silicon burning and form a core composed mostly of iron-group elements. Since these are the most stable elements, no more nuclear energy can be extracted to counter-balance gravity. Once the inner core is massive enough for electron degeneracy pressure to be overcome (Chandrasekhar mass limit), the iron core collapses
and sometimes leads to a powerful explosion as in the case of SN1987A.
In this paper, we review the general evolution of massive stars at solar metallicity as well as the impact of rotation and mass loss. We also discuss briefly the effects of metallicity. The models and plots 
presented in this paper are taken from \citet{psn04a} unless otherwise stated. Other recent grids of models at solar metallicity can be found in \citet{grids12, 2013ApJ...764...21C,2014ApJ...783...10S}. 

\section{Evolution of Surface Properties (HR diagram) and Lifetimes}
\label{ltau}
\begin{figure}[!tbp]
\centering
   \includegraphics[width=0.5\textwidth]{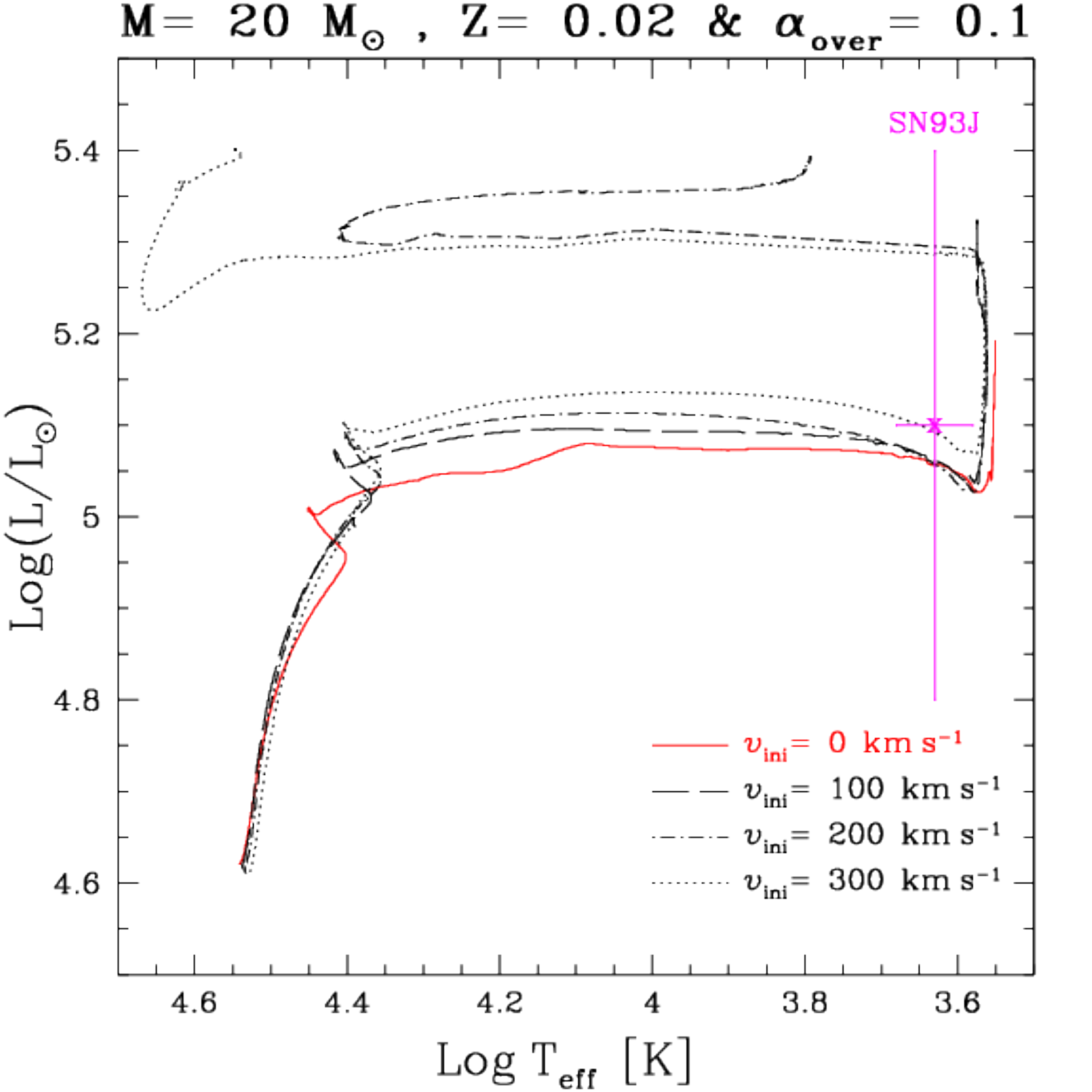}\includegraphics[width=0.5\textwidth]{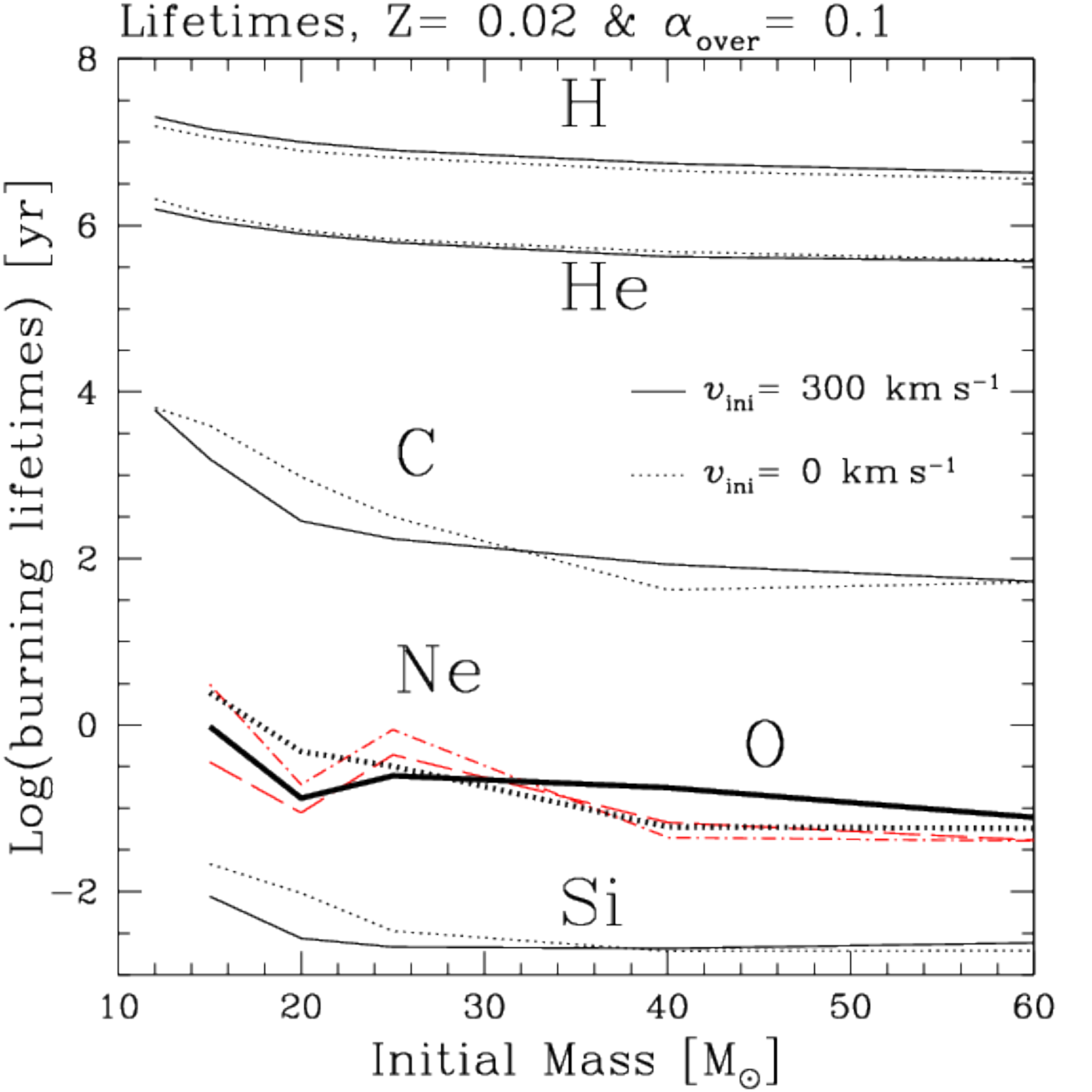}

\caption{{\it Left}: HR--diagram for 20 $M_{\odot}$ models: solid, dashed, dotted-dashed and dotted lines correspond 
respectively to $v_{\rm{ini}}$= 0, 100, 200 and 300 km\,s$^{-1}$.
We also indicate the position of the progenitor of SN1993J. {\it Right}: Burning lifetimes as a function of the initial 
mass and 
velocity. Solid and dotted lines correspond respectively to rotating and
non--rotating models. Long--dashed and dotted--dashed lines are used for
rotating and non--rotating Ne--burning lifetimes to point out that they
are to be considered as estimates.}
\label{hr20}
\end{figure}
Figure \ref{hr20} ({\it left}) shows the evolutionary tracks of different 20\,$M_{\odot}$ models  in  the  HR--diagram 
and thus how the surface properties of these stars evolve.
The non-rotating model is representative of the lower end of massive stars, which keep an extended hydrogen-rich 
envelope, end as red supergiant and produce type II supernovae. The 300 km\,s$^{-1}$ model is representative of the 
higher end of massive stars, for which most or all of the H-rich envelope is lost via stellar winds and the star ends 
as a hot star, generally a Wolf-Rayet star and produce a type Ib or Ic supernova depending on how much helium is left. 
Note that type Ib and Ic supernovae also come from stars in multiple systems, in which the hydrogen-rich envelope is lost via Roche lobe overflow \citep[see][for a review on the topic of binary interactions in massive stars]{l12}.
These two models also show the impact of rotation on the evolution of massive stars. 
The additional models with intermediate rotation ($v_{\rm{ini}}$= 100 and 200 km\,s$^{-1}$) show the smooth transition from non-rotating to fast rotating models \citep[for HR diagrams covering the full IMF, see][]{grids12}.

As mentioned above, massive stars go through 6 burning stages: H, He, C, Ne, O, and Si burning.   
The lifetimes of these stages are plotted in Fig. \ref{hr20} ({\it right}). Whereas H and He-burning stages last 
for roughly 10$^{6-7}$ and 10$^{5-6}$ years, respectively, the lifetimes for the advanced phases is much shorter. This 
is due to neutrino losses dominating energy losses over radiation from C burning onwards. C, Ne, O ans Si burning 
phases last about 10$^{2-3}$, 1, 1 and 10$^{-2}$ years respectively.
Concerning the effects of rotation and mass loss, there is a mass range where rotational mixing 
($M \lesssim 30 M_{\odot}$) or
mass loss ($M \gtrsim 30 M_{\odot}$) dominates over the other process. For $M \lesssim 30 M_{\odot}$, rotation-induced 
mixing extends the H-burning lifetime and as a consequence shortens slightly He-burning lifetimes. For the advanced 
phases, rotation makes stars behave like more massive stars. This is clearly seen for C-burning lifetimes, which are 
shorter for rotating models. For $M \gtrsim 30 M_{\odot}$, strong mass loss leads to degeneracy in the lifetime and 
final properties.

\begin{figure*}[!tbp]
\centering
\includegraphics[width=0.44\textwidth]{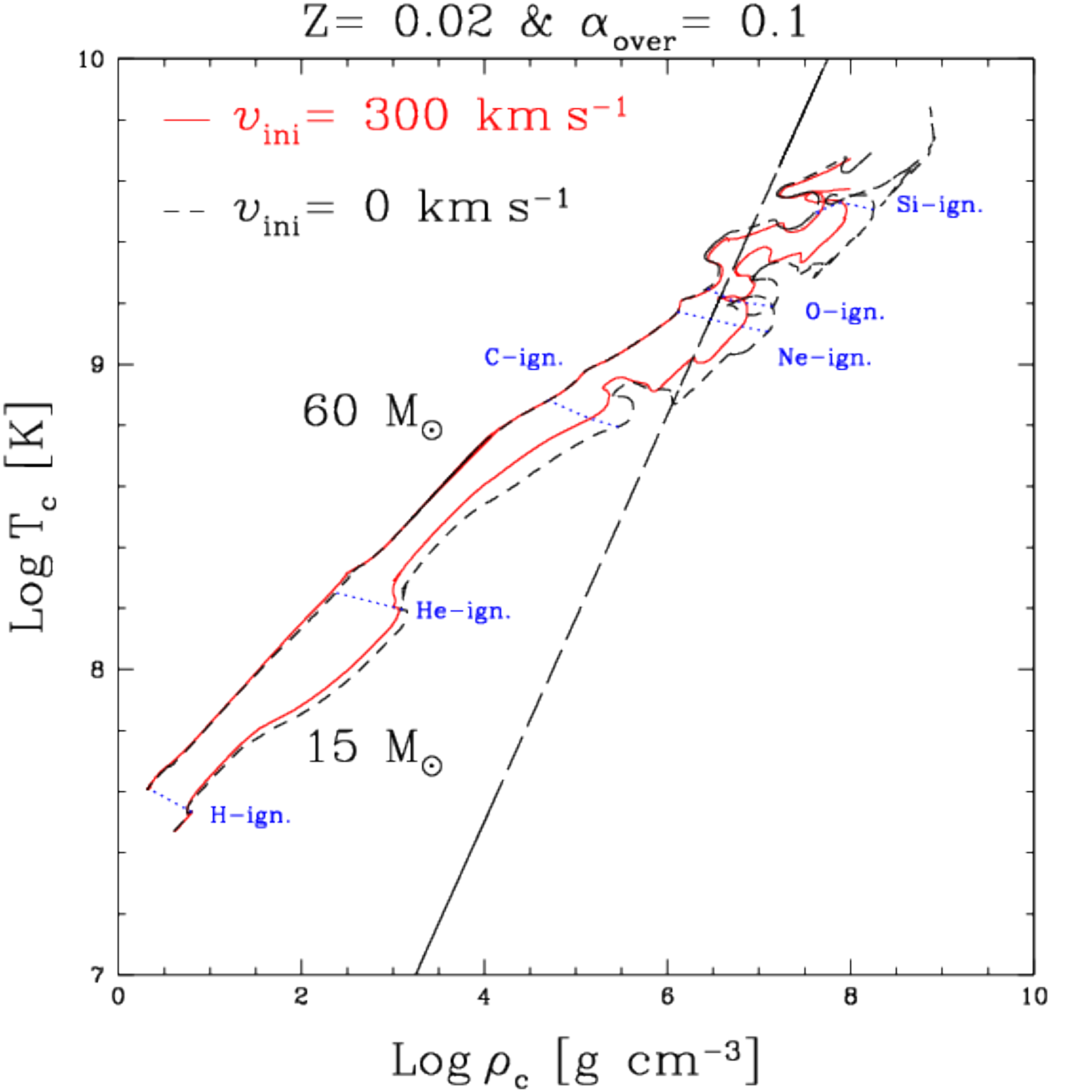}\includegraphics[width=0.3\textheight,angle=90]{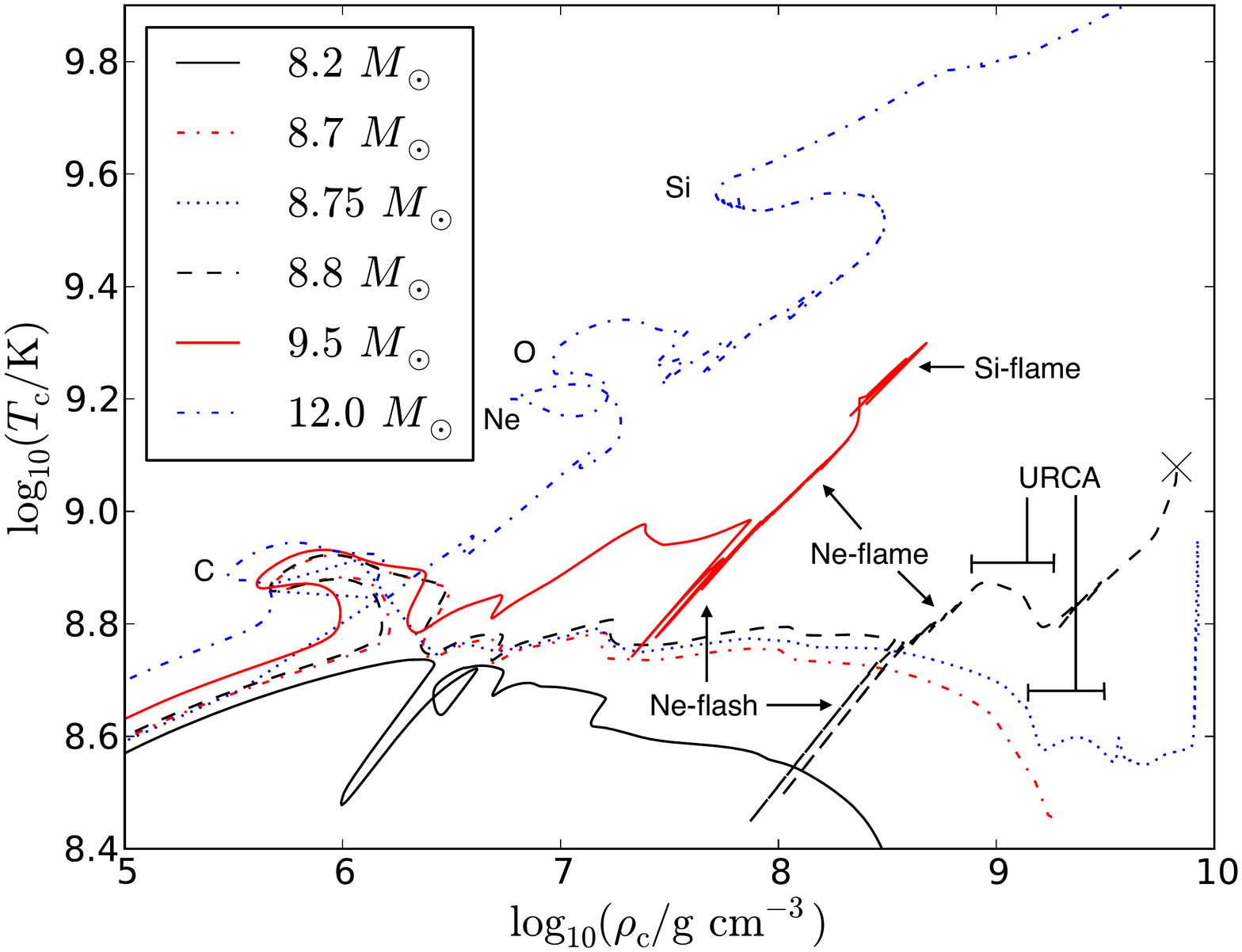}
\caption{Log\,$T_{\rm{c}}$  vs Log\,$\rho_{\rm{c}}$ diagrams: 
Left: evolutionary tracks for the 15 and 60 $M_{\odot}$ models.
Solid lines are rotating models and dashed lines are non--rotating models. 
The ignition points of every burning stage are connected with dotted lines.
The additional long dashed line
corresponds  to the limit between non--degenerate and degenerate
electron
gas ($P^{\rm{el}}_{\rm{perfect\,gas}}=P^{\rm{el}}_{\rm{degenerate\,gas}}$). 
 Figure taken from \citet{psn04a}.
Right: The divergence of the models following C-burning 
at the transition between massive and intermediate mass stars. Figure taken from \citet{2013ApJ...772..150J}.
}
\label{trsl}
\end{figure*}

\section{Evolution of Central Properties in the Log\,$T_{\rm{c}}$--Log\,$\rho_{\rm{c}}$ Diagram and Lower Mass Limit for Massive Stars}
Figure \ref{trsl} (left) shows the tracks of the 
15 and 60 $M_{\odot}$ models
throughout their evolution 
in
the central temperature versus central
density plane (Log\,$T_{\rm{c}}$--Log\,$\rho_{\rm{c}}$ diagram). 
The 60 $M_{\odot}$ model is representative of the stars more massive than about 30 $M_{\odot}$, for which the evolution is mainly affected by mass loss. 
The 15 $M_{\odot}$ model is representative of the stars less massive  than about 20 $M_{\odot}$, for which the evolution
is mainly affected by rotational mixing (already identified in Sect. \ref{ltau}).
For the 15 $M_{\odot}$ model, the rotating 
tracks have a higher temperature and lower density due to more massive convective cores.
The  bigger cores are due to the effect of mixing, which largely
dominates the structural effects of the centrifugal force.
On the other hand, for the 60 $M_{\odot}$ model, mass loss dominates
mixing effects and the rotating model tracks in the 
Log\,$T_{\rm{c}}$--Log\,$\rho_{\rm{c}}$ plane
are at the same level or
below the non--rotating ones \citep[see][for more details]{psn04a}.

Recent models for the transition between massive and intermediate stars ($8-12\,M_\odot$ stars) can be found in 
\citet{2013ApJ...772..150J,2013ApJ...771...28T,2015ApJ...810...34W} and older models can be found in 
\citep{Nomoto1984,Nomoto1987,Ritossa1999}. Evolutionary tracks from 
\citet{2013ApJ...772..150J} are shown in Fig.\,\ref{trsl}. Similar trends and conclusions are found in the 
other studies. The evolution and fate of stars in this mass range are sensitive to convective boundary mixing (CBM) 
treatment (e.g. overshooting), mass loss and CO core growth. Different choices of CBM lead to the transition mass being 
shifted up and down but we expect the same transitions and regimes to take place for different choices of CBM. The fate 
of super-AGB stars (SAGB, AGB stars that undergo carbon burning but not neon or subsequent burning stages) is highly 
sensitive to the mass--loss prescription on the SAGB and the rate at which the core grows \citep{Poelarends2008}. Mass 
loss and core growth compete against each other. 
At solar metallicity mass loss often wins and only a very narrow mass range at the 
top of the SAGB mass range will end as {electron-capture supernovae} (ECSN). The 8.7 and $8.75\,M_\odot$ models represent 
models in this narrow mass range. The $8.2\,M_\odot$ model represents models in the SAGB mass range for which mass 
loss wins and this model will end as a ONe white dwarf (WD). 
A temperature inversion develops in the core following the extinction of carbon-burning in both the $8.8\,M_\odot$ and 
$9.5\,M_\odot$ models. The neutrino emission processes that remove energy from the core are (over-)compensated by 
heating from gravitational contraction in more massive stars. However in these lower-mass stars the onset of partial 
degeneracy moderates the rate of contraction and hence neutrino losses dominate, cooling the central region. As a 
result, the ignition of neon in the 8.8 and 9.5~$M_\odot$ models takes place off center. These two models then go 
through neon(/oxygen; oxygen also burns via fusion in this situation)-flashes followed by the development of a 
neon(/oxygen)-flame. Owing to the high densities in the cores of these stars, the products of neon and oxygen burning 
are more neutron-rich than in more massive stars. This results in an electron fraction in the shell of as low as 
$Y_\mathrm{e}\approx0.48$. Due to its higher degeneracy, $Y_\mathrm{e}$ decreases faster in the $8.8\,M_\odot$ and it 
contracts faster than the time needed for the neon/oxygen flame to reach the centre of the star, both processes being 
helped by URCA pair processes. The core of the $8.8\,M_\odot$ model continuously contracts until the center reaches
the critical density for electron captures by $^{24}\mathrm{Mg}$, quickly
followed by further contraction to the critical density for those by
$^{20}\mathrm{Ne}$ and this model results in core 
collapse. The $8.8\,M_\odot$ model produces a ECSNe as for the $8.75\,M_\odot$ model but via 
a new evolutionary path coined ``failed massive star'' by \citet{2013ApJ...772..150J} rather than via the SAGB evolutionary path. The ``failed massive 
star'' path is also expected to take place for a narrow mass range but it does not critically depend on the 
uncertainties linked to mass loss, which is the case for the SAGB progenitors of ECSNe.
Similarly to the $8.8\,M_\odot$ model starting neon burning off centre, the $9.5\,M_\odot$ model starts silicon burning 
off centre in a shell that later propagates toward the center. This is another example of the continuous transition 
towards massive stars, in which all the burning stages begin centrally. Although this model was not evolved to its 
conclusion, we expect that silicon-burning will migrate to the center, producing an iron core, and that it will finally 
collapse as an iron core-collapse SN (FeCCSN). The canonical massive star evolution (igniting C-, Ne-, O- and 
Si-burning centrally) leading to FeCCSN is expected to take place for stars with masses above $10\,M_\odot$. This mass 
range is represented by the $12\,M_\odot$. 

The fate of models at the lower end of massive stars is studied further in \citet{2014ApJ...797...83J,2015ApJ...810...34W,2016A&A...593A..72J}.
At the other end of the IMF, very massive stars evolve far away from degeneracy. Very massive stars may encounter 
instead the pair-creation instability at very high temperatures \citep[see][and references therein]{VMSbook}.


\section{Structure Evolution and Pre-Supernova Properties} \label{presn}
\begin{figure*}[!htp]
\centering
\includegraphics[width=0.5\textwidth]{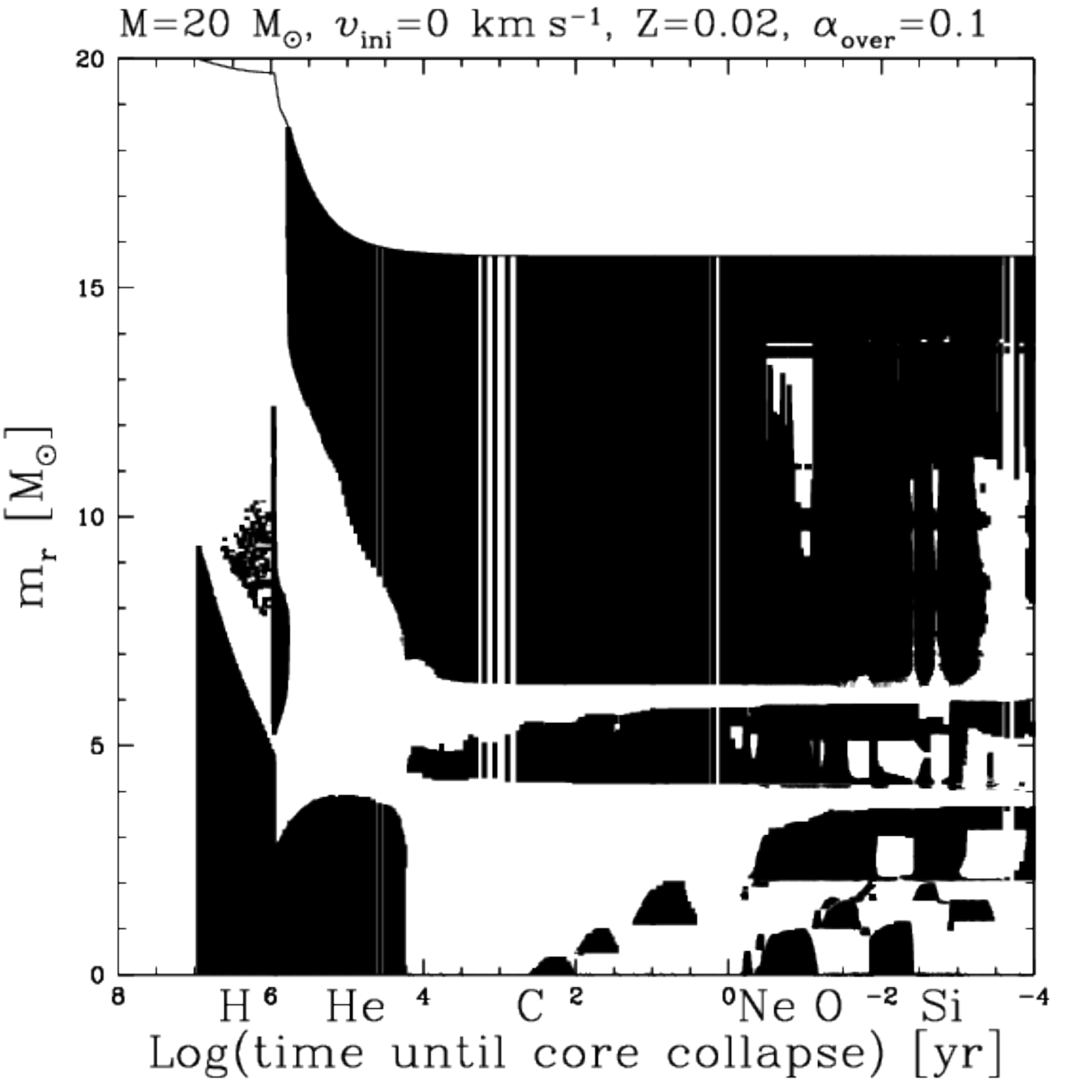}\includegraphics[width=0.5\textwidth]{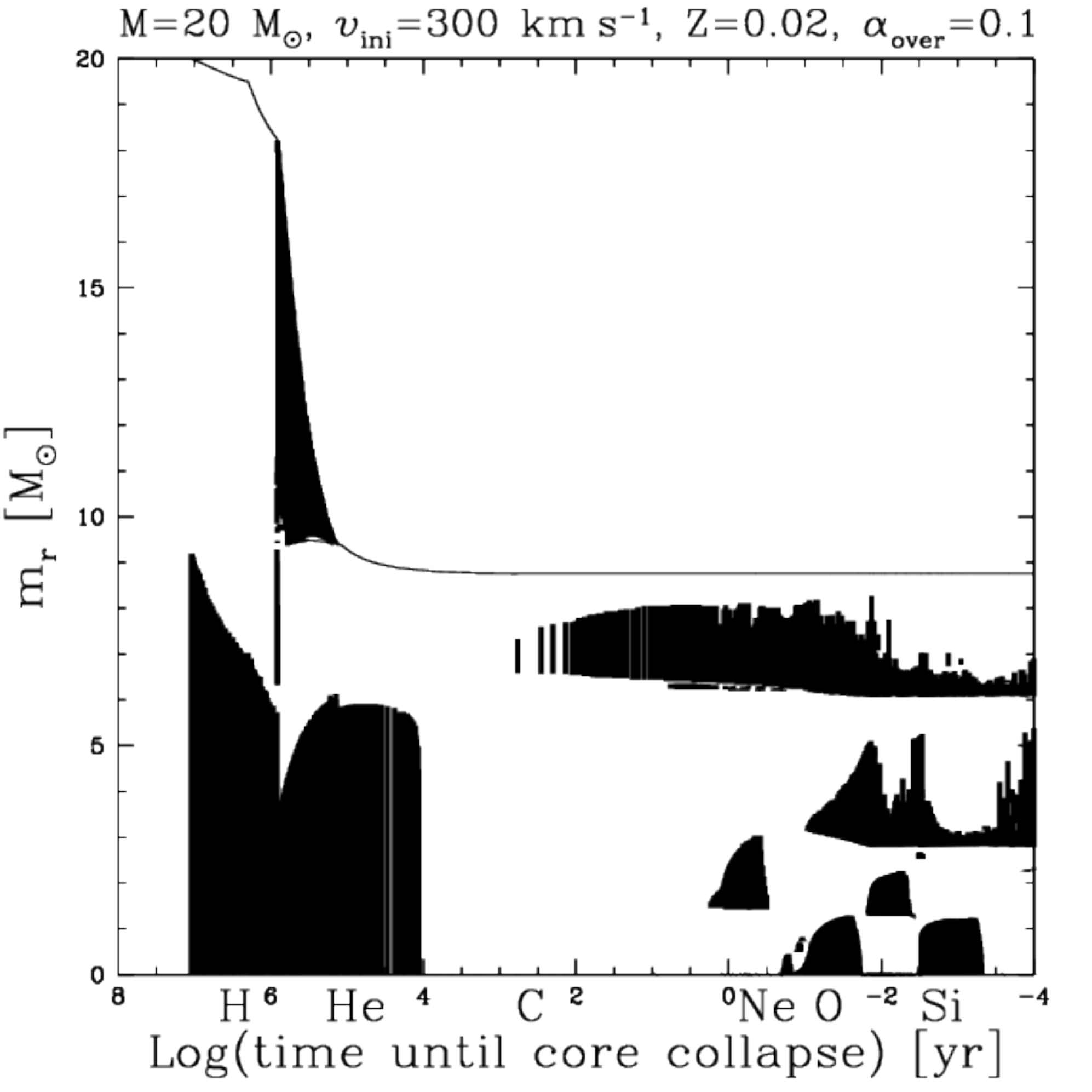}
\caption{Kippenhahn diagrams for the non--rotating ({\it left}) and 
$v_{\rm{ini}}$= 300 km\,s$^{-1}$ ({\it right}) 20 $M_{\odot}$ models. 
The black zones correspond to convective regions. Note that these plots are produced by drawing 
black vertical lines for a subset of the time steps of the model and thus the vertical white lines around log(time left 
until collapse) $\sim 3$ are only due to the drawing technique and the models remain convective in between neighbouring 
black vertical lines.
}
\label{dhr5m121520}
\end{figure*}

Figure \ref{dhr5m121520} shows the evolution of the structure (Kippenhahn diagram) for 20 $M_{\odot}$ models. 
The y--axis represents the mass 
coordinate and the x--axis 
the time left until core collapse. The black zones represent convective zones.
The abbreviations of the various burning stages are written below the
graph at the time corresponding to the central burning stages.
We note the complex succession of the different convective zones during the advanced phases. 
\citet{2014ApJ...783...10S} study in detail the complex convective history in massive stars. In particular, their fig. 
13 shows how the location in mass of the lower boundary of carbon burning convective shells play a key role in 
determining the compactness \citep{2011ApJ...730...70O} at the {pre-supernova} stage \citep[see also][]{2016ApJ...818..124E}. It is worth noting that a few 
physical ingredients of the stellar models influence carbon burning in general and thus the exact location of the 
convective shells and the compactness for a given initial mass. Carbon burning is sensitive to the amount of carbon 
(relative to oxygen) left at the end of helium burning. This in turn is influenced by the 
$^{12}$C($\alpha,\gamma$)$^{16}$O rate relative to the triple-$\alpha$ rate \citep{2009ApJ...702.1068T}. Convective 
boundary criteria and mixing prescriptions also affect the carbon left over at the end of helium burning. Using Ledoux 
rather than Schwarzschild generally leads to smaller helium burning cores and more carbon left over. Extra mixing, 
especially towards the end of He-burning brings fresh $\alpha$ particles that can capture on $^{12}$C and reduce 
its left over abundance. Finally rotation-induced mixing, as is clearly seen in Fig.\,\ref{dhr5m121520}, leads to 
significantly larger helium cores, less left over carbon and leads to radiative core carbon burning ({\it right} 
panel). 
The other differences between
non-rotating and rotating models are the following. We can see that small convective zones above the central H-burning
core disappear in rotating models. Also visible is the loss of the
hydrogen rich envelope in the rotating models. The non-rotating 20 $M_{\odot}$ model is 
representative of the stars below 20 $M_\odot$, while the rotating 20 $M_{\odot}$ model is 
representative of the non--rotating and rotating models above 30 $M_{\odot}$. Above 30 $M_{\odot}$,
all stars have very similar convective history after He--burning. They all lose their H-rich envelope and 
undergo core C burning under radiative conditions. The main difference between stars above 30 $M_{\odot}$ and the 
rotating 20 $M_{\odot}$ model is that stars above 30 $M_{\odot}$ have one large carbon convective shell that sits 
around 3 $M_{\odot}$ and thus does not influence much the final stages and the compactness at the pre-supernova stages. 
The complex history of convective zones and the uncertainties in the input physics mentioned here make it very hard to 
predict the exact explosion properties of a star of a given initial mass. Nevertheless, it is likely, as in the case of SAGB 
stars, that the same transitions would occur (e.\,g. from convective to radiative core carbon burning) even if the 
input physics changes. Convective boundary mixing (CBM) during carbon burning (and other stages), 
if able to change the extent of convective burning 
shells may affect the compactness of supernova progenitor significantly. 3D hydrodynamic simulations of convective 
boundary mixing will hopefully help constrain the 1D prescriptions used in stellar 
evolution codes \citep[see][for a review on the topic]{2015ApJ...809...30A}. We present in Sect. \ref{3d} new 3D simulations of a carbon-burning shell in a 15 $M_{\odot}$ model \citep{cshell}, which brings new light on CBM during this stage. 

\begin{figure*}[!htbp]
\centering
   \includegraphics[width=0.33\textwidth]{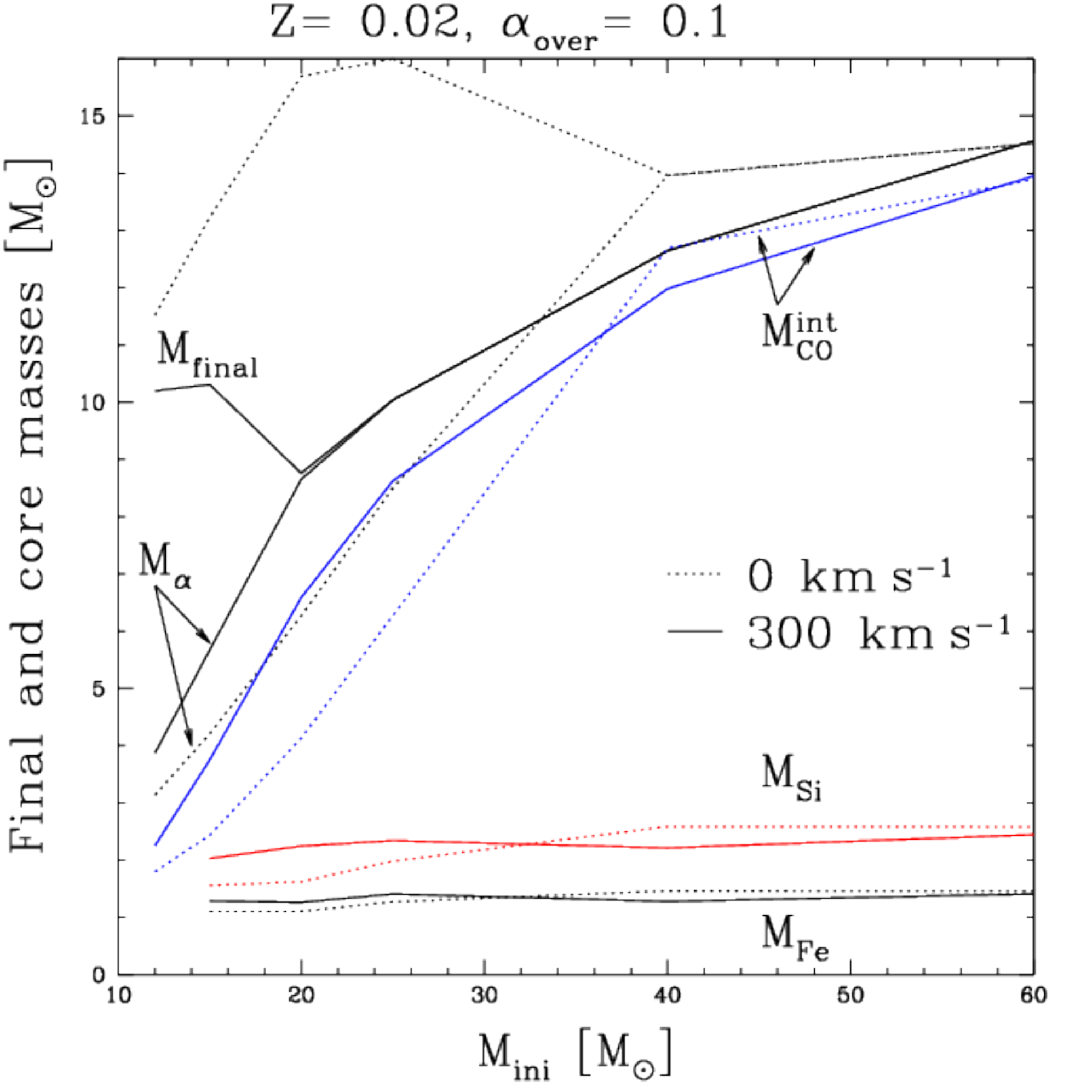}\includegraphics[width=0.33\textwidth]{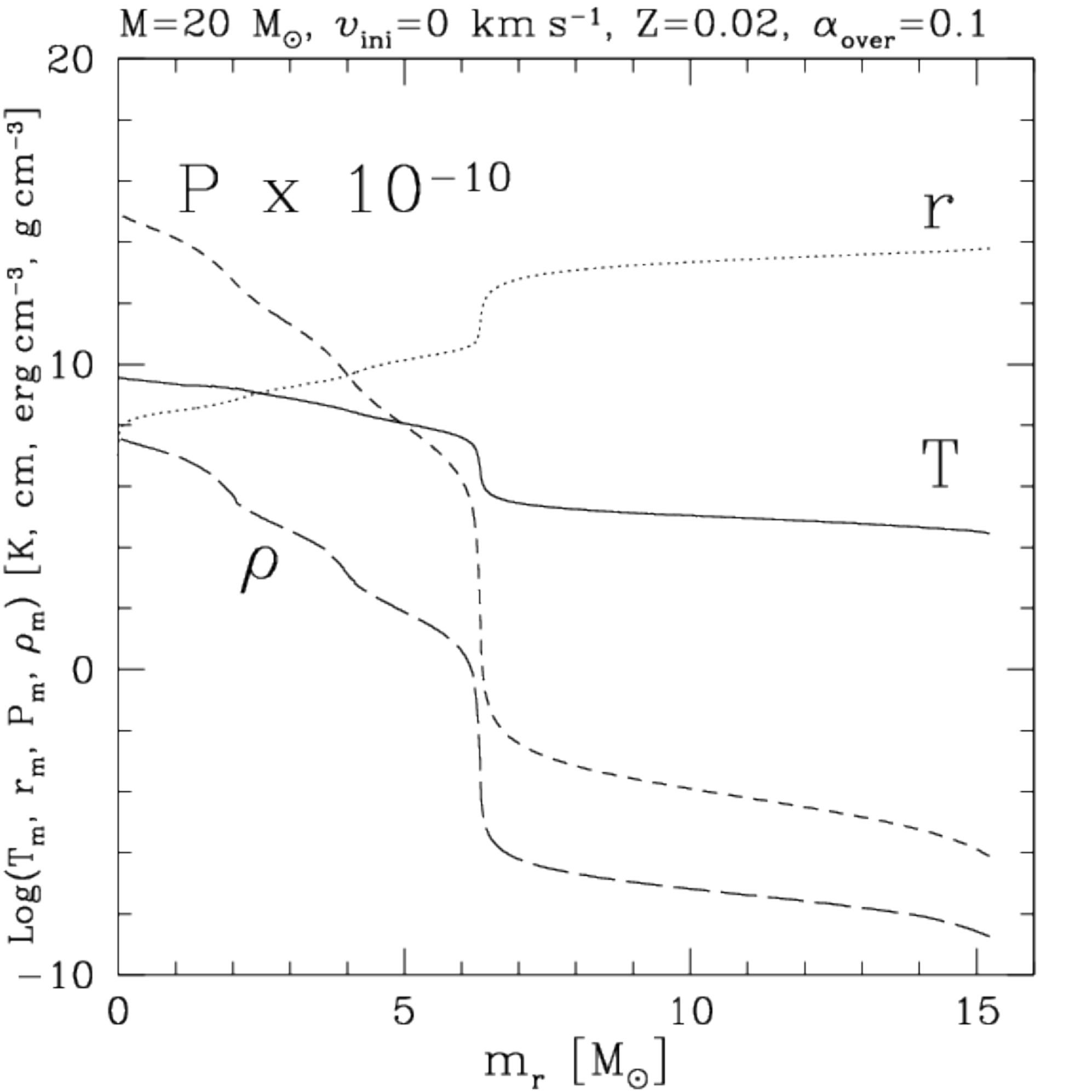}\includegraphics[width=0.33\textwidth]{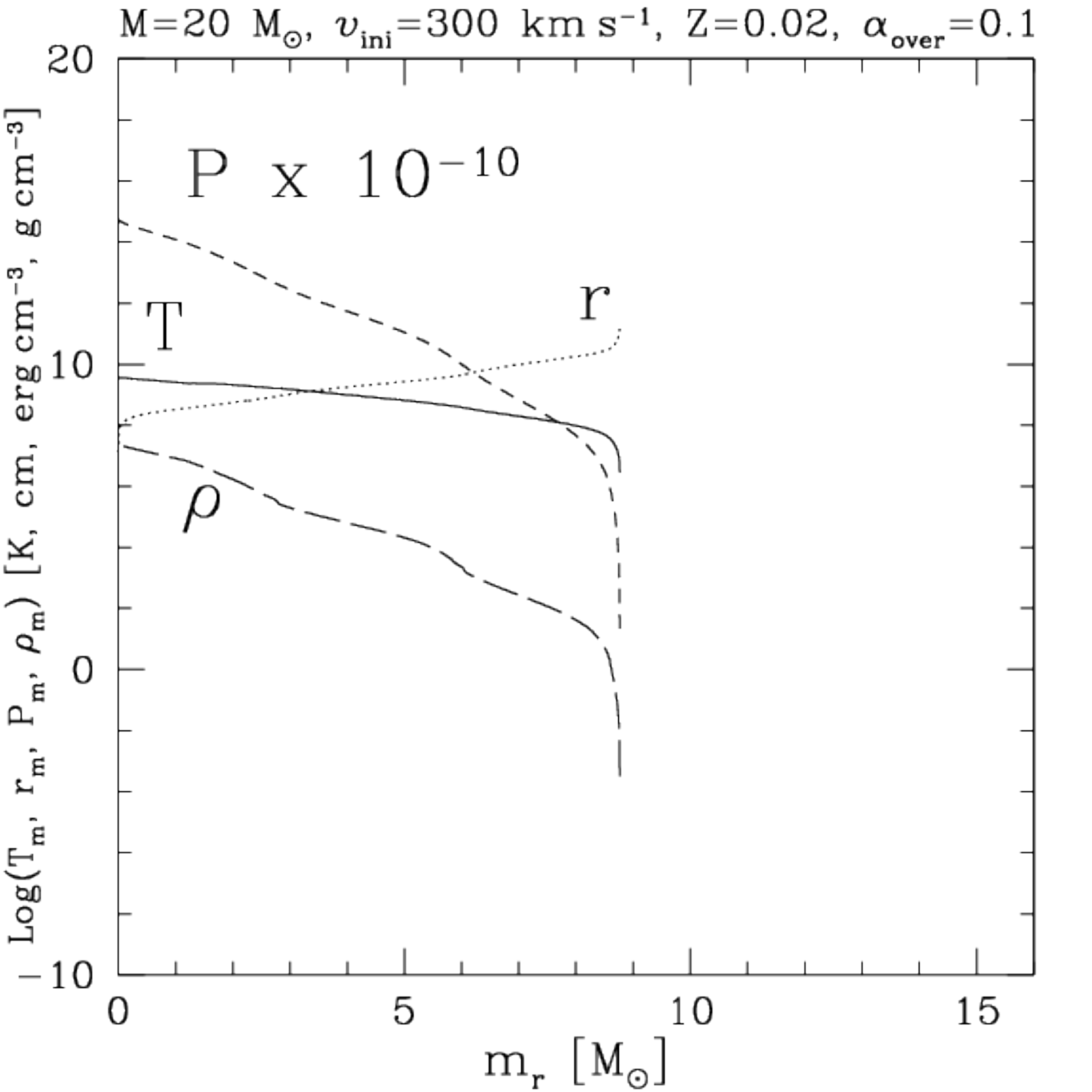}
\caption{Pre-supernova properties (end of Si-burning): core masses as a function of the initial mass and velocity ({\it Left}). 
Profiles of the radius, $r$, density, $\rho$, temperature, $T$ and pressure $P$ for the
non--rotating ({\it middle}) and rotating ({\it right}) 20 $M_{\odot}$ models. The
pressure has been divided by $10^{10}$ to fit it in the diagram.
}
\label{steq}
\end{figure*}

Figure \ref{steq} ({\it left}) shows the core masses 
as a function of initial mass
for non--rotating (dotted lines) and rotating  (solid lines) models. 
Since rotation increases mass loss, the final
mass, $M_{\rm{final}}$, of rotating models is always smaller than that of non--rotating
ones. Note that for stars with $M \gtrsim 50 M_{\odot}$ mass
loss during the WR phase is proportional to the actual mass
of the star. This produces a convergence of the final masses \citep[see for instance][]{ROTXI}.
We can again see a general difference between the effects of rotation
below and above $30\,M_{\odot}$.
For $M \lesssim 30\,M_{\odot}$, rotation significantly increases 
the core masses due to mixing.  
For $M \gtrsim 30\,M_{\odot}$, rotation makes the star enter the WR phase at an earlier
stage. The rotating star spends therefore a longer time in this
phase characterised by heavy mass loss rates. This results in smaller
cores at the pre-supernova stage.
We can see in Fig.~\ref{steq} that the difference between rotating and
non--rotating models is the largest between 15 and 25 $M_{\odot}$. As explained above, this will have an impact on the 
compactness in this sensitive mass range. Improvements in input physics may reconcile model 
predictions with observationally determined masses of type II supernova, with a maximum below 20 $M_{\odot}$, named the 
RSG problem by \citet{2009ARA&A..47...63S} if the mass range of high compactness ends up covering the mass range between 
about 17 and 22 $M_{\odot}$, while more massive stars explode as type Ib or Ic supernova or fail to explode \citep[see also][]{wr12, 2012MNRAS.419.2054W}.


As well as the chemical composition (abundance profiles and core masses) 
at the pre-supernova stage, other properties, like the density profile, 
the neutron excess, the entropy and 
the total radius of the star, play an important role in the supernova
explosion. Figure \ref{steq} ({\it middle}: non-rotating  and {\it right}: rotating 20 $M_{\odot}$ models) shows the density, temperature, radius and pressure
variations as a function of the Lagrangian mass coordinate at the end
of the core Si--burning phase. The
rotating model has lost its envelope, this truly affects the parameters
towards the surface of the star. The radius of the star (BSG) is about
one percent that of the non--rotating star (RSG). As said above this modifies
strongly the supernova explosion. We also see
that temperature, density and pressure profiles are flatter 
in the interior of rotating models due to the bigger core sizes. 

\section{Metallicity Effects}
\label{mdot}
The effects of {metallicity} on stellar evolution are described in several studies 
\citep[see for example][]{HFWLH03,CL04,MM94}.
A lower metallicity implies a lower luminosity which leads to slightly smaller
convective cores.  
A lower metallicity also implies lower opacity and lower
mass losses (as long as the chemical composition has not been changed by
burning or mixing in the part of the star one considers). So at the start
of the evolution lower metallicity stars are more compact and thus have bluer tracks during the main
sequence.
The lower metallicity models also have a harder time reaching
the red supergiant (RSG) stage 
\citep[see][ for a detailed discussion]{ROTVII}. Non--rotating models 
around $Z=10^{-3}$
becomes a RSG only after the end of core He-burning and
lower metallicity non--rotating models never reach the RSG stage. At even lower metallicities, as long as the
metallicity is above about $Z=10^{-10}$, no significant differences have
been found in non-rotating models. Below this metallicity and for metal
free stars, the CNO cycle cannot operate at the start of H--burning. At
the end of its formation, the
star therefore contracts until it starts He-burning because the
pp--chains cannot balance the effect of the gravitational force. Once
enough carbon and oxygen are produced, the CNO cycle can operate and the
star behaves like stars with $Z>10^{-10}$ for the rest of the main
sequence. Shell H--burning still differs between $Z>10^{-10}$ and metal
free stars. Metal free stellar models are presented in \citet{CL04},
\citet{UN05}, \citet{EM08} and \citet{2010ApJ...724..341H}.

How does rotation change this picture? At all metallicities, rotation 
usually increases the core sizes, the lifetimes, the
luminosity and the mass loss.
\citet{ROTVII} and \citet{ROTVIII} show that rotation 
favours a redward evolution and that rotating models better reproduce the
observed ratio of blue to red supergiants (B/R) in the small Magellanic cloud.
Rotating models
around $Z=10^{-5}$ become RSGs during shell He--burning. 
This does not
change the ratio B/R but changes the structure of the star when the SN
explodes.
At even lower metallicities \citep[$Z=10^{-8}$ models presented in][]{H07}, the 20 $M_\odot$ models do not become RSG. 
However, more massive models do reach the RSG stage and the 85 $M_\odot$ model even
becomes a WR star of type WO (see below).
\citet{ROTVII} also find that a larger fraction of stars reach break-up velocities
during the evolution. The impact of rotation on nucleosynthesis and in particular the (not-so) weak $s$-process in low-$Z$ massive rotating stars was studied in detail by \citet{sproc12,sproc16}.

Note that fast rotation at low metallicities may lead to energetic explosions such 
as gamma-ray bursts \citep[see e.g.][]{HMM05,HWS05,YLN06}.

\section{3D Hydrodynamic Simulations of Convection in Carbon Burning}\label{3d}
Due to the complex nature of stars, stellar models would ideally require three-dimensional (3D) (magneto-)hydrodynamic models that include all the relevant physics. 3D hydro models must use time steps that are at most days. The total lifetime of stars, however, is at least millions of years. This explains why most stellar evolution models are limited to (spherically-symmetric) one dimension (1D). The predictive power of 1D models, however, is crippled by 1D prescriptions of 3D phenomena containing free parameters, which need to be tuned to reproduce subsets of observations. A key uncertain prescription in 1D codes is that of convection, in particular convective boundary mixing (CBM). The mixing length theory (MLT) of convection used in most codes dates back to \citet{1958ZA.....46..108B}. The main deficiency of MLT \citep[or MLT updates, e.\,g.][]{1991ApJ...370..295C} is that it does not provide a treatment of the convective boundary. For the boundary placement, codes must use even simpler prescriptions based on linear analysis (Schwarzschild or Ledoux criterion) and {\it have to} add CBM to reproduce the main-sequence (MS) width \citep{BMC11,grids12}. The post-MS evolution is strongly affected by these choices, leading to uncertain predictions \citep{2013A&A...560A..16M}. Furthermore, 
SN progenitor structure is sensitive to the convection history as mentioned in Sect. \ref{presn}.
Asteroseismic observations are able to constrain further CBM choices  \citep{2014A&A...572L...5M,2014MNRAS.439L...6G}. 
However, additional guidance and constraints are needed to improve CBM prescriptions in 1D codes and in turn the predictive power of stellar evolution.
The computing power has finally reached the point where convective boundaries can be resolved in the largest simulations with $\ge 1024^3$  resolution \citep[see e.g.][]{2015ApJ...798...49W,2015ApJ...809...30A, 2017MNRAS.465.2991J}.
This means that now these simulations can provide the guidance and constraints to build the next generation of stellar evolution models including 3D-hydro-based  ($+$asteroseismology-guided) prescriptions for CBM. 

\begin{figure}
\begin{center}
 \includegraphics[width=0.33\linewidth]{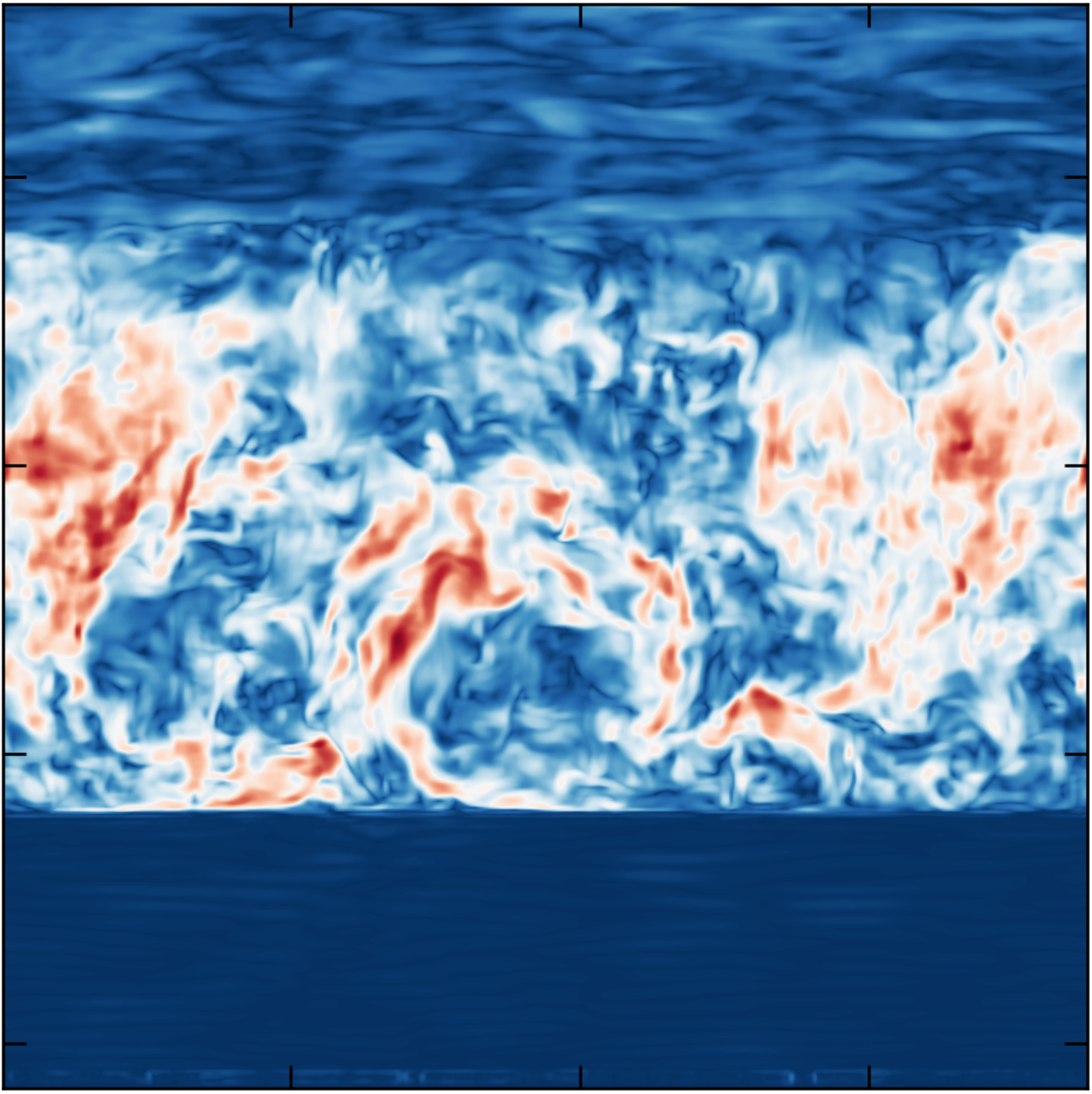}\includegraphics[height=0.33\linewidth]{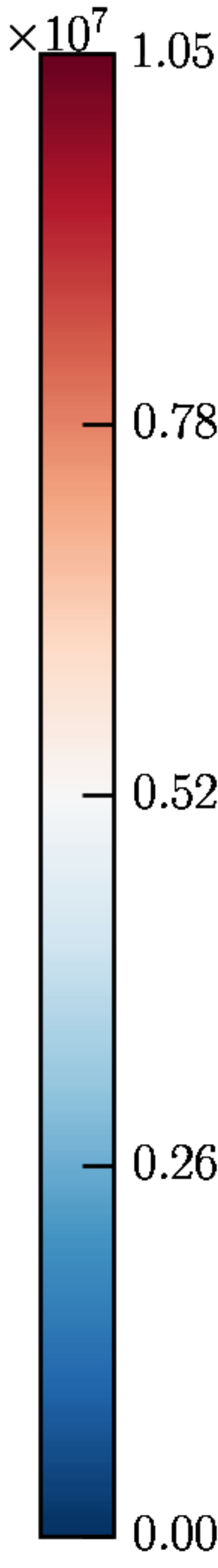}\includegraphics[width=0.66\textwidth]{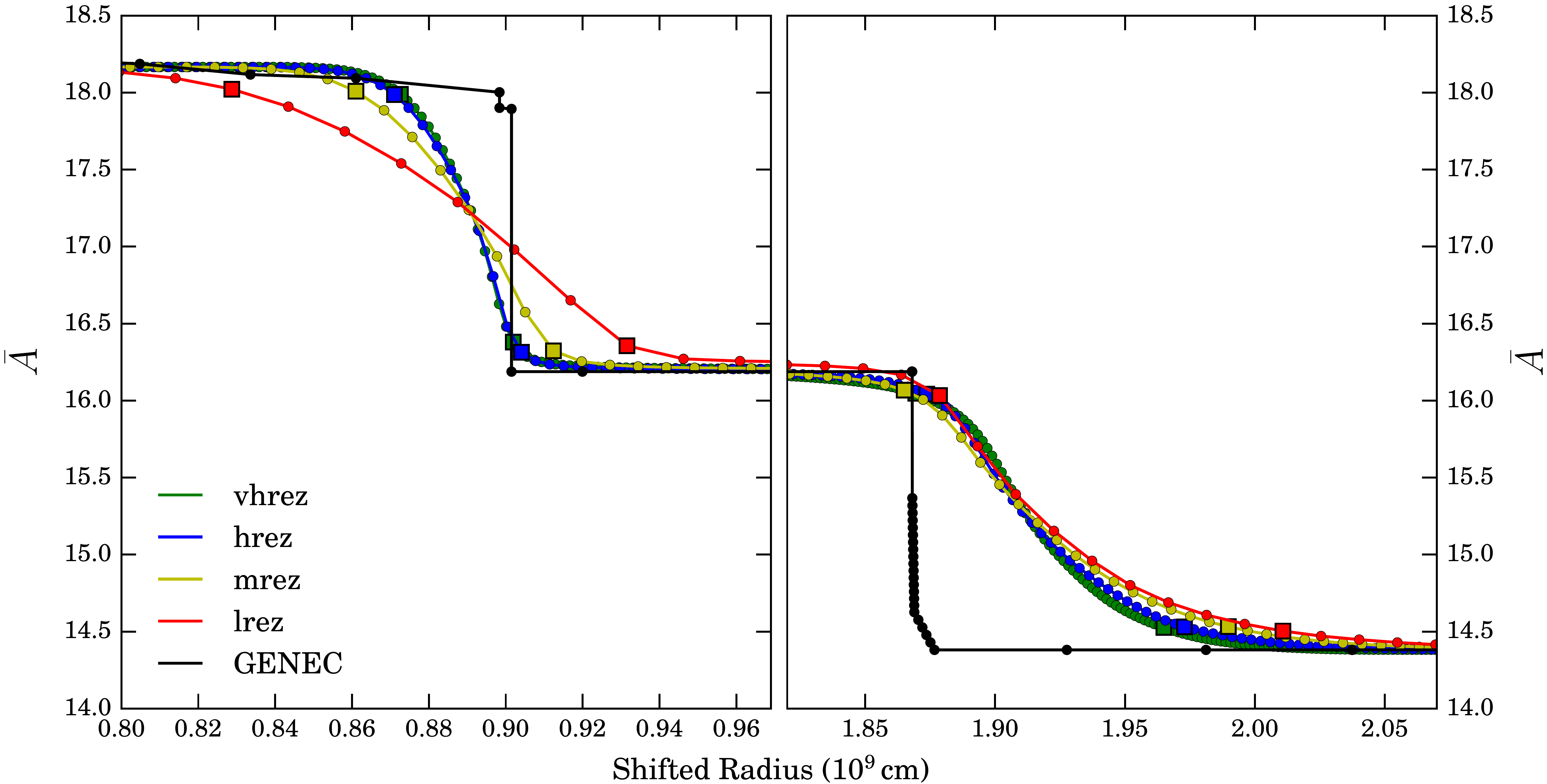}
\end{center}
 \caption{Vertical (2D) snapshot taken at 2,820\,s into the 512$^3$ (\textsf{hrez}) simulation of the velocity magnitude in the plane of the snapshot, $\sqrt{v_x^2 + v_y^2}$, $x$ being the vertical direction ({\it left}). The colour-map represents the velocity magnitude in cm\,s$^{-1}$. Radial compositional profiles at the lower ({\it middle}) and upper ({\it right}) convective boundary regions for the last time step of each model. 
The radius of each profile is shifted such that the boundary position coincides with the boundary position of the \textsf{vhrez} model. 
In this sense, it is easier to assess the convergence of each model's representation of the boundary at the final time-step. Individual 
mesh points are denoted by filled circles. Approximate boundary extent (width) is indicated by the distance between two filled squares for 
each resolution. The initial composition profile provided by the 1D  
\textsc{genec} code \citep{psn04a} is shown in black for a qualitative comparison only \citep[see][for more details]{cshell}.}
\label{abar_cz}
\end{figure} 

We summarise in this section the results of \citet{cshell} who studied CBM in a carbon-burning convective shell in a 15 $M_\odot$ model using the \textsc{prompi} code \citep{2007ApJ...667..448M}. A snapshot from these simulations is shown in Fig.\ref{abar_cz} ({\it left}). 
Entrainment of material at both convective boundaries was observed in all the simulations and was analysed in the framework of the entrainment law \citep{2004JAtS...61..281F,2014AMS...1935.JGJG}. 
The entrainment rate in our simulations was found to be roughly inversely proportional to the 
bulk Richardson number, Ri$_{\rm B}$ ($\propto $Ri$^{-\alpha}$, $0.5\lesssim \alpha \lesssim 1.0$). 
We also found that convective boundaries are broadened by shear mixing (Kelvin-Helmholtz instability) due to the fact that the flow has to do a U-turn at the boundaries. We estimated the boundary widths (see Fig. \ref{abar_cz} {\it middle} and {\it right}) and found these to be roughly 30\% and 10\% of the local pressure scale height for 
the upper and lower convective boundary, respectively. While these widths are only estimates, they confirm that the 
lower boundary is narrower than the upper boundary. More importantly, the abundance profiles in the 3D simulations is much smoother than in the 1D model, which does not include any CBM.

\section{Summary and Outlook}
In this paper, we reviewed the general properties and evolution of CCSNe progenitors in the framework of 1D stellar evolution models. We also discussed the limitations of 1D models and how 3D hydrodynamic simulations can provide unique and crucial constraints for convective boundary mixing in massive stars, which is one of the major uncertainties affecting the progenitors of CCSNe. We then presented new 3D hydrodynamic simulations of convection in carbon burning. 
These new carbon burning simulations confirm the general findings of the oxygen shell simulations presented in \citet{2007ApJ...667..448M}. This is promising for the long term goal of developing a convective boundary mixing prescription for 1D models 
which is applicable to all (or many) stages of the evolution of stars (and not only to the specific conditions studied in 3D simulations). The 
luminosity (driving convection) and the bulk Richardson number (a measure of the boundary stiffness) will  
be key quantities for such new prescriptions \citep[also see][]{2015ApJ...809...30A}. 

The goal of 1D stellar evolution models is to capture the long-term (secular) evolution of the convective zones and of 
their boundaries, while 3D hydrodynamic simulations probe the short-term (dynamical) evolution. Keeping this in mind, the 
key points to take from existing 3D hydrodynamic studies for the development of new prescriptions in 1D 
stellar evolution codes are the following:
\begin{itemize}
 \item Entrainment of the boundary and mixing across it occurs both at the top and bottom boundaries. Thus 1D 
stellar evolution models should include convective boundary mixing at both boundaries. Furthermore, the boundary shape 
is not a discontinuity in the 3D hydrodynamic simulations but a smooth function of radius, sigmoid-like, a feature that 
should also be incorporated in 1D models.
 \item At the lower boundary, which is stiffer, the entrainment is slower and the boundary width is narrower. This 
confirms the dependence of entrainment and mixing on the stiffness of the boundary. 
 \item Since the boundary stiffness varies both in time and with the convective boundary considered, a single constant 
parameter is probably not going to correctly represent the dependence of the mixing on the instantaneous convective 
boundary properties. As discussed above, we suggest the 
use of the bulk Richardson number in new prescriptions to include this dependence. 
More in-depth discussions on this topic can be found in \citet{2015ApJ...809...30A,2015A&A...580A..61V}
\end{itemize}

\section*{Acknowledgements}
The authors acknowledge support from EU-FP7-ERC-2012-St Grant 306901. RH acknowledges support from the World Premier International Research Centre Initiative (WPI Initiative), MEXT, Japan and the “ChETEC” COST Action (CA16117), supported by COST (European Cooperation in Science and Technology). 
This work used 
the Extreme Science and Engineering Discovery Environment (XSEDE), which is supported by National Science Foundation grant number OCI-1053575. CM and WDA acknowledge support from NSF grant 1107445 at the 
University of Arizona. The authors acknowledge the Texas Advanced Computing Center (TACC) at The University of Texas at Austin (http://www.tacc.utexas.edu) for providing HPC resources that have contributed 
to the research results reported within this paper. This work used the DiRAC Data Centric system at 
Durham University, operated by the Institute for Computational Cosmology on behalf of the STFC DiRAC HPC Facility (www.dirac.ac.uk). This equipment was funded by BIS National E-infrastructure capital grant 
ST/K00042X/1, STFC capital grants ST/H008519/1 and ST/K00087X/1, STFC DiRAC Operations grant ST/K003267/1 and Durham University. DiRAC is part of the National E-Infrastructure.

\bibliographystyle{aa}

\end{document}